# The tunneling spectra and superconducting gaps observed by using scanning tunneling microscope near the grain-boundary of $FeSe_{0.3}Te_{0.7}$ films

K.C. Lin[1,2], Y.S. Li[2], Y.T. Shen[1], M.K. Wu[2,3], and C.C. Chi[1]

[1] *Department of Physics, National Tsing Hua University, Hsinchu 300, Taiwan*
[2] *Institute of Physics, Academia Sinica, Nankang, Taipei 115, Taiwan*
[3] *Department of Physics, National Dong Hwa University, Hualien 974, Taiwan*

**ABSTRACT**

We used STM to study the tunneling spectra of $FeSe_{0.3}Te_{0.7}$ films with two orientations of ab-planes and the connection ramp between them. We have discovered that, using pulse laser deposition (PLD) method, the a- and b-axis of the $FeSe_{0.3}Te_{0.7}$ film deposited on Ar-ion-milled Magnesium Oxide (MgO) substrate are rotated 45° with respect to those of MgO, while the a- and b-axis of the film grown on pristine MgO substrate are parallel to those of MgO. With photolithography and this technique, we can prepare $FeSe_{0.3}Te_{0.7}$ film with two kinds of orientations on the same MgO substrate, and the connection between them forms a ramp at an angle about 25° to the substrate plane. In the planar region with either 0° or 45° orientation, we have observed tunneling spectra with a superconducting gap about 5 meV and 1.78 meV respectively. However, a much larger gap about 18 meV has been observed in the ramp region. Furthermore, we have also observed a small zero-bias conductance peak (ZBCP) inside the large gap at T = 4.3K. The ZBCP becomes smaller and disappear with increasing temperature to 7 K. The possible causes of such large gap and the ZBCP will be discussed.

## 1. INTRODUCTION

The new series of iron-based superconductor in fluorine-doped LaFeAsO with critical temperature ($T_C$) at 26 K was discovered by Hideo Hosono and co-workers in February 2008 [1]. This discovery triggers the beginning of world-wide efforts to study the new iron-based superconductors. Several months later, the iron-chalcogenide superconductor (FeSe) with $T_C$ of 8 K was reported by Hsu et al. [2]. Up to now, the iron-based superconductor can be roughly divided into five series [3-7], and among these series of superconductors, the FeSe has the simplest structure. By partially substituting Se with Tellurium (Te), which has a larger ion radius, can enhance $T_C$ up to ~14K [8, 9]. Furthermore, $T_C$ of FeSe can also be raised to 37K by applying high pressure of 8.9 GPa [10], and, to a lesser degree, it can also be enhanced by epitaxial strain [11-15].

In this paper, we report the scanning tunneling microscope (STM) and scanning tunneling spectroscopy (STS) study on the oxygen-annealing FeSe$_{0.3}$Te$_{0.7}$ (FeSeTe) thin film deposited on Magnesium Oxide (MgO) substrate with different orientations by pulse laser deposition technique (PLD). The critical temperatures ($T_C$) of the two orientations are around 12.5 K in non-rotated plane and 12 K in rotated plane [16]. Despite of the fact that the superconducting transition temperatures of the two differently oriented grains are rather close, the superconducting energy gaps, determined by STS, differs by nearly a factor of 3. Even more surprisingly, we have observed a very large gap of 18 meV in the boundary region between the two grains. Furthermore, we have also observed a zero-bias conductance peak (ZBCP) in the boundary region at 4.3 K.

## 2. EXPERIMENT PROCEDURE

Similar to the case of depositing YBa$_2$Cu$_3$O$_{7-\delta}$ on YSZ substrate [17, 18], we have discovered that, using pulse laser deposition (PLD) method, the a- and b-axis of the FeSe$_{0.3}$Te$_{0.7}$ film deposited on Ar-ion-milled Magnesium Oxide (MgO (100)) substrate are rotated 45° with respect to those of MgO, while the a- and b-axis of the film grown on pristine MgO substrate are parallel to those of MgO. With photolithography and this technique, we can prepare FeSe$_{0.3}$Te$_{0.7}$ film with two kinds of orientations on the same MgO substrate, and the connection between them forms a ramp at an angle about 25° to the substrate plane [19]. The sample fabrication process is as follows: First, the 20nm of FeSeTe film was deposited on pristine MgO substrate in a vacuum chamber (<5×10$^{-5}$ mbar). Next, we defined the striped pattern of FeSeTe thin films with 10 μm wide on MgO substrate by the standard photo-lithography technique. By using the Argon ion-milling process, the striped pattern with a ramp of angle about 25° with respect to substrate plane can be formed. After gently cleaning the sample by low-voltage Argon ion-milling, 100nm FeSeTe film was immediately deposited on the stripe films. Finally, the sample was annealed in 1 bar O$_2$ for 12 hours to enhance the film critical temperature ($T_C$) [16]. Actually, the $T_C$ of non-rotated and rotated plane is around 12.5 K and 12 K respectively by our previous experiments [16]. The in-plane φ-scans of X-ray diffractometer (XRD), as shown in Figure 1, indicate that the (202) crystalline-axis of FeSeTe film grown on the virgin MgO substrate is parallel to the (101) crystalline-axis of MgO substrate, but the same (202) crystalline-axis of FeSeTe film grown on the bombarded MgO substrate is

45°-rotated in the ab-plane with respect to the (101) MgO substrate. The schematic diagram of the sample and its orientation is shown in Fig. 2.

After the sample fabrication, we put the sample into a high vacuum load-lock chamber, which is connected to the ultrahigh vacuum (UHV) chamber of the STM system, to have the sample surface cleaned by an Ar-ion milling process for 15 min at 500 V to reduce the surface oxide layer, and followed by another 15 min at 300 V to smoothen the surface. Finally, the sample is annealed at 280°C for 30 min to repair the possible surface damage by ion-milling.

The STM used in this study is a commercial low temperature ultrahigh-vacuum (LT-UHV) STM apparatus designed and manufactured to our specification by the Unisoku Co. of Japan. Prior to taking the tunneling data, a Pt-Ir tip was cleaned in-situ by field emission to ensure no loose particles or molecules on the tip. The tunneling conductance spectra were acquired by using a standard lock-in amplifier with a small bias voltage modulation of 0.3 mV at 337 Hz superimposed on a dc bias voltage. The STM is used to image the surface topography. And the scanning tunneling spectroscopy (STS) is used to probe the quasiparticle density of states (DOS) and to measure the superconducting gap at the Fermi energy.

3. RESULTS AND DISCUSSIONS

Figure 3(a) shows a typical STM topography across a grain boundary, taken at T = 4.3 K with a tunneling current of 50 pA and a set bias voltage of 30 mV in a scanning range about 400 nm. The STM topographic image reveals a clear ramp on the surface. The line scan (black line in Fig. 3(a)) in Fig. 3(b) shows that the step height is about 25 nm and the ramp angle with respect to the substrate plane is about 25°. Figure 3(c) and 3(d) are the STM topographic images of the non-rotated surface (on the higher plane side of the ramp) and the rotated surface (on the lower plane of the ramp) respectively. It is clear that the orientations of the grains, indicated by the black arrows are consistent with the directions of their crystalline axes. We note that the grain size, ~50 nm$^2$, on the non-rotated plane is larger than the grain size, ~20 nm$^2$, on the rotated plane. We believe that the difference of grain size is related to the surface defects present on the substrate. To make sure that the grain orientations are not influenced by the large-angle grain boundaries in the ramp, the images were taken in a region at least 100 nm away from the ramp.

Eight points evenly spaced at 50 nm separation, indicated by arrows with letters from a to h along the line scan shown Fig. 3(b), are the chosen locations to analyze the evolution in superconducting gap with positions from their corresponding tunneling conductance spectra shown in Figure 4. In Fig. 4(a), we found that the voltage of the peak, taken from the left-hand side of the conductance curve, varies from 5 mV to 18 mV corresponding to position a to d. However, when conductance spectra were taken from the bottom of ramp to the rotated plane (position e to h), the coherence peaks disappear, and the U-shape curve becomes a V-shape curve when the conductance spectra are taken on the rotated plane. It is worth nothing that the conductance spectrum taken at position d, on the ramp but close to the rotated plane, shows a ZBCP at 4.3 K. Although the ZBCP of FeSeTe superconductor has never been reported in other STM literatures [21, 22, 23, 24], our other transport experiments including the in-plane and c-axis tunneling junction have also observed this anomaly [16, 19]. Hence, to better understand the superconducting gap and ZBCP evolution with temperature, we take three points, on the non-rotated plane (position a), on the boundary ramp (position d), and on the rotated plane (position h) to measure the evolution of the tunneling spectra with temperature, shown in Figure 4(b), 4(c) and 4(d) respectively. From Fig. 4(b), the coherence peak, usually associated with the superconducting gap at about 5.2 meV at 4.3 K, reduces and broadens substantially as temperature increases from 4.3 K to 9 K. In contrast to the usual BCS superconductors, such as Nb, the reduced temperature for the coherence peak to disappear is way too low. Thus we cannot reliably plot the gap versus temperature curve. Figure 4(c) shows a much larger gap structure of 18.6 meV with clear coherence peaks at 4.3K. We defer the discussion of such large gap in a later session. Probably due to the size of the gap, the coherence peaks survive up to 9K. The temperature dependence of the gap does not seem to be monotonic. It is interesting to note that the small ZBCP clearly observable at 4.3 K quickly diminishes with increasing temperature and finally disappears at 6 K. Figure 4(d) shows no coherence peak at any temperatures when the conductance spectra were taken far away from boundary at position h. It is not unusual to observe no clear coherence peaks in the STM tunneling conductance spectra for FeSeTe thin films [21, 23, 24]. In this situation, one usually associate the turning point, such as the one at 1.78 mV at T = 4.3 K in Fig. 4(d), to be the superconducting gap.

The changes of the conductance spectrum from position a to h, shown in Fig. 4(a), are quite surprising and unexpected. The coherence peaks become much larger from position a to d and the gap size increases from 5.2 meV to 18.6 meV. To our knowledge, such a large gap for Fe-(11) series has only been reported once by Q.Y. Wang et al. [20]. They observed two superconducting gaps of 20.1 meV and 9 meV at

4.2K with a $T_C$ of 53 K on a one-unit-cell FeSe film grown in-situ with a MBE system equipped with an attached LT-STM probe. The authors claim that such a large gap may be caused by some interface effects between the film and substrate or by stress on this film due to the lattice-mismatched STO substrate. Actually, the enhanced superconducting transition temperatures of FeSe and FeSeTe have been observed with applied pressure [10] or the epitaxial strain [11-15]. The STM topography results of Fig. 2(c) and 2(d) suggest that a large stress can be caused in the large-angle grain boundary region due to the push and pull from each side of the grains so that a larger gap is induced. The potential higher $T_C$ in this narrow region might be suppressed by the proximity effect of the two lower $T_C$ planes. That might also explain the non-BCS behavior observed for our sample.

## 4. CONCLUSION

In summary, we have successfully fabricated epitaxial $FeSe_{0.3}Te_{0.7}$ films on MgO substrate with its in-plane crystalline axis either parallel to or rotated 45° with respect to the MgO lattice. We have observed a superconducting gap about ~5.2 meV in the non-rotated plane and 1.78 meV in the rotated plane. The values of the superconducting gaps are close to some reported results of FeSeTe [16, 19, 21, 22, 23]. Besides, a much larger gap at about ~18.6 meV and a small ZBCP is observed in the boundary region at T = 4.3 K. We suggest that the existence of larger gap may result from the strain on the boundary and as a result, the superconductivity will be enhanced. We speculate that the ZBCP structure may be caused by impurity or magnetic field trapped on boundary.

## ACKNOWLEDGEMENT

This work was supported by National Science Council, Taiwan, R.O.C. Grant Number: NSC 101-2112-M-007-013.

**FIGURE CAPTIONS**

**FIG. 1.** (a) The in-plane ϕ-scan of the FeSeTe film deposited on the pristine MgO substrate shows that the in-plane crystalline-axes of the FeSeTe film is parallel to those of MgO. (b) The in-plane ϕ-scan of the FeSeTe film deposited on the Ar-ion bombarded MgO substrate shows that the in-plane crystalline-axis of FeSeTe film is rotated 45° with respect to those of MgO substrate. The diffraction peaks used for the ϕ-scans of the FeSeTe film and the MgO substrate are (202) and (101) respectively.

**FIG. 2.** A schematic diagram of the cut-up side view of the sample is shown in (a). The gradient color on the ramp indicates that the orientation of FeSeTe grain growth may vary along with the ramp. The black concave area on substrate indicates the substrate have been bombarded by low-voltage Argon ions prior to the film deposition. (b) A schematic diagram of the top view of the sample with arrows indicating the in-plane crystalline axes of the FeSeTe film.

**FIG. 3.** (a) A typical STM topography, across a 45° grain-boundary ramp, was taken in a constant-current scanning mode with a tunneling current of 50 pA and a bias voltage of 30 mV. (b) Line scan image (black line in (a)). The *dI/dV* spectra shown in Figure 4 are taken at the positions indicated by the arrowheads with symbols a to h. (c) STM topography of non-rotated surface (on the higher plane) taken far away from ramp for clear image. The grain size is around ~50 $nm^2$. The arrowhead (black arrow) indicates the direction of grain growth. (d) STM topography of rotated surface (on a lower plane) taken on opposite direction away from ramp. The grain size is around ~20 $nm^2$. The angle between these two directions of grain growth is about 43°.

**FIG. 4.** Four series of *dI/dV* spectra were taken with sample bias voltage of 30 mV and tunneling current of 50 pA. Bias-modulation amplitude was set at 0.3 mV at 337 Hz. (a) Tunneling *dI/dV* spectra taken at 4.3K along the line indicated in Fig. 3(a) at equal separation (50 nm). We found that the voltage of the peak in the left-hand side varies from 5 mV to 18 mV when the tip moves from position a to d. The coherence peaks are absent for positions e to h. (b) The *dI/dV* spectra taken at position a with temperature varied from 4.3 K to 9 K. There is a clear gap about 5.2 meV at 4.3 K. The coherence peaks disappear when temperature is higher than 6K. (c) The *dI/dV* spectra taken at position d with temperature varied from 4.3 K to 9 K. We observed a clear large gap feature around 18.6 meV and a small ZBCP at 4.3 K. The coherence peaks disappear when temperature is higher than 8 K. (d) The *dI/dV* spectra taken at position h with temperature varied from 4.3 K to 9 K. There is a turning point located

about 1.78 mV at 4.3K, which is usually associated with the superconducting gap.

**Fig. 1**

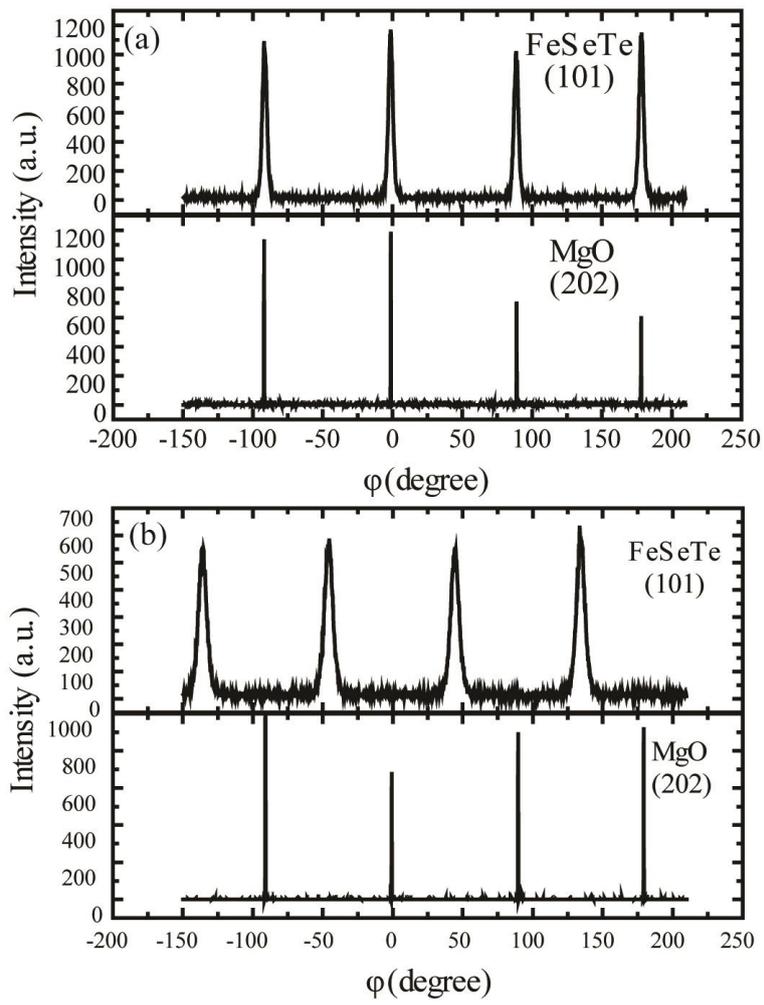

**Fig. 2**

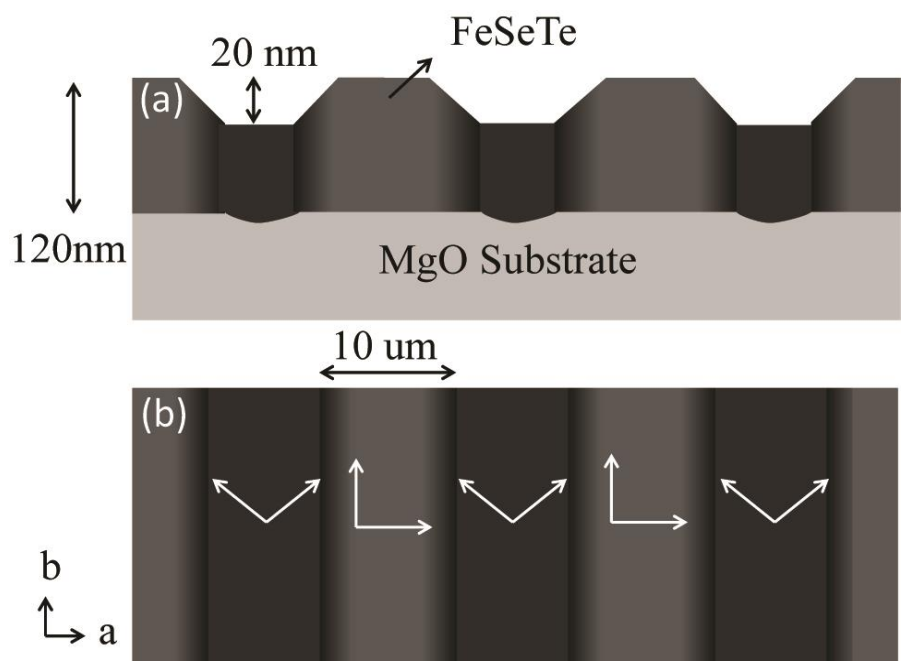

**Fig. 3**

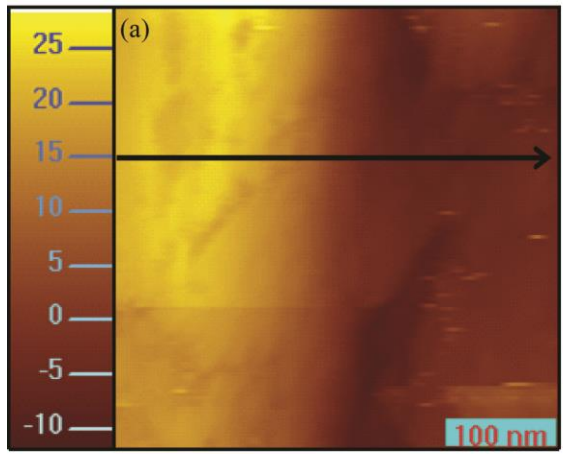 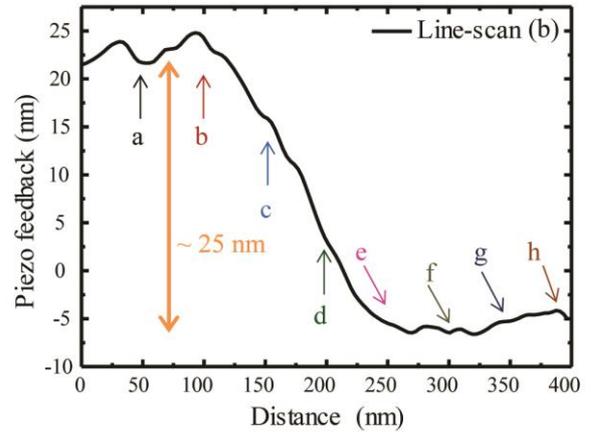

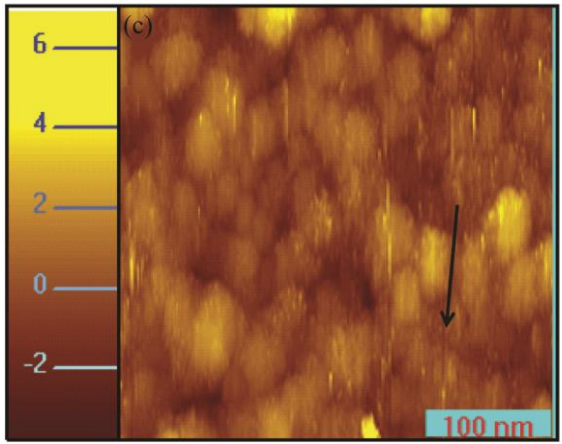 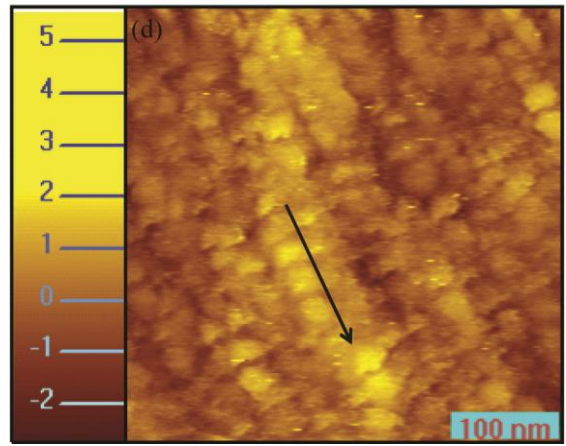

**Fig. 4**

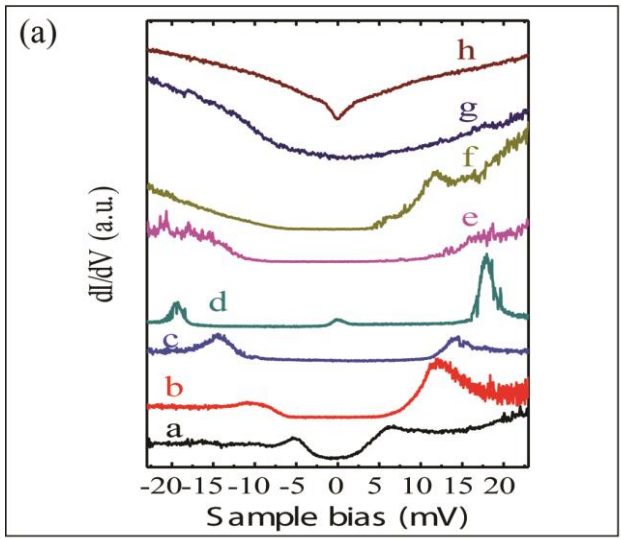
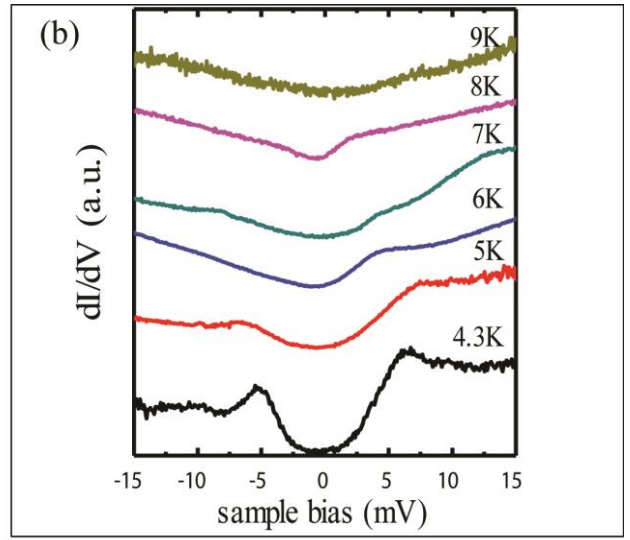
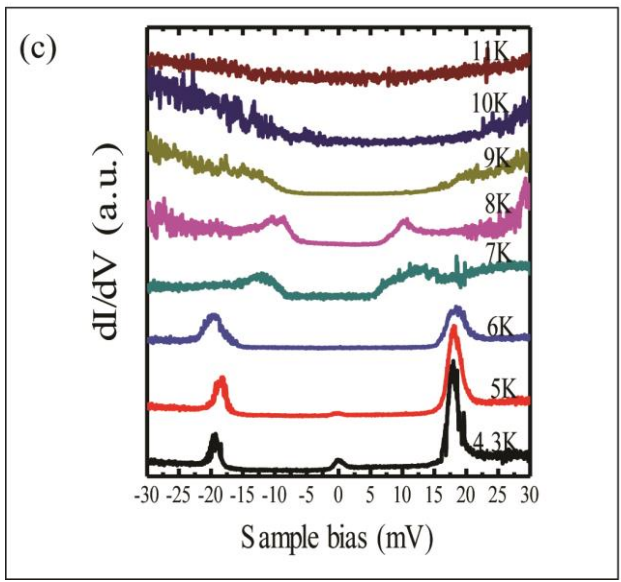
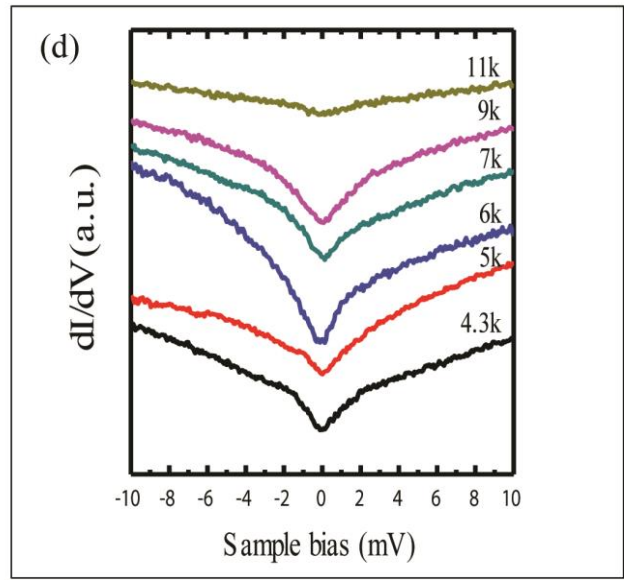